\documentclass[11pt,epsbox]{article}
\usepackage{amsmath,amssymb}
\usepackage{graphicx}
\usepackage{enumerate}

\baselineskip=\normalbaselineskip
\renewcommand{\baselinestretch}{1.4}
\makeatletter
\@addtoreset{equation}{section}

\makeatother
\newcommand{\be}{\begin{equation}}
\newcommand{\ee}{\end{equation}}
\newcommand{\ba}{\begin{eqnarray}}
\newcommand{\ea}{\end{eqnarray}}

\newcommand{\del}{\partial}
\newcommand{\bra}[1]{\left\langle\,{#1}\,\right|}
\newcommand{\ket}[1]{\left|\,{#1}\,\right\rangle}
\newcommand{\bracket}[2]{
\left\langle\left.\,{#1}\,\right|\,{#2}\,\right\rangle}

\newcommand{\bz}{\overline{z}}

\newcommand{\hO}{\hat{O}}

\newcommand{\hPsi}{\hat{\Psi}}

\newcommand{\hN}{\hat{N}}

\newcommand{\Tr}{{\rm Tr}}

\newcommand{\st}{\star}

\newcommand{\hU}{\hat{U} }

\newcommand{\daga}{\hat{a}^{\dagger}}

\begin{document}
\setcounter{page}{0}

\begin{flushright}
\parbox{40mm}{%
Preprint TU-912 \\
June 2012
}
\end{flushright}

\vfill

\begin{center}
{\Large{\bf 
Angles in Fuzzy Disc  and  Angular Noncommutative Solitons
}}
\end{center}

\vfill

\renewcommand{\baselinestretch}{1.0}

\begin{center}
\textsc{Shinpei Kobayashi}
\footnote{E-mail: \texttt{shimpei@nat.gunma-ct.ac.jp}} 
and
\textsc{Tsuguhiko Asakawa}
\footnote{E-mail: \texttt{asakawa@tuhep.phys.tohoku.ac.jp}} 


~\\
$^1$ \textsl{Department of Physics, 
      Gunma National College of Technology,  \\
     580 Toribamachi, Maebashi, 371-8530, JAPAN}  

\vspace{0.5cm}

$^{2}$ \textsl{Department of Physics, Graduate School of Science, \\
Tohoku University, 
Sendai 980-8578, JAPAN} \\ 

\end{center}

\begin{center}
{\bf abstract}
\end{center}

\begin{quote}

\small{%
The fuzzy disc, introduced by the authors of \cite{Lizzi:2003ru},
is a disc-shaped region in a noncommutative plane,
and is a fuzzy approximation of a commutative disc.
In this paper we show that one can introduce
a concept of angles to the  fuzzy disc,
by using the phase operator and phase states known in quantum optics.
We gave a description of the fuzzy disc in terms of operators and
their commutation relations,
and studied properties of angular projection operators.
A similar construction for the fuzzy annulus is also given.
As an application, we constructed fan-shaped soliton solutions
of a scalar field theory on the fuzzy disc.
We also applied this concept to the theory of noncommutative
gravity we proposed in \cite{Asakawa:2009yb}.
In addition, possible connections to some systems in physics
are suggested.}
\end{quote}
\vfill

\renewcommand{\baselinestretch}{1.4}

\renewcommand{\thefootnote}{\arabic{footnote}}
\setcounter{footnote}{0}
\addtocounter{page}{1}
\newpage

\tableofcontents

\section{Introduction}
\label{sec:intro}

Noncommutative geometry and its applications to field theories 
have been extensively investigated for a long time. 
Naive motivation for them would come from how we should 
quantize spacetime. Although there is no reliable answer for 
this question, the concept of quantum geometry and 
the physics related to it seem very fascinating. 

It is known that there are nontrivial solutions for a theory 
on such a noncommutative space in spite that 
it does not have any nontrivial solution in its commutative limit. 
The GMS soliton, which is the solution of a scalar 
field theory on a three-dimensional spacetime with 
noncommutativity between spatial coordinates,  
is a  well-known example for that \cite{Gopakumar:2000zd}. 
There the noncommutativity is introduced by 
the ``canonical" commutation relation $[\hat{x}, \hat{y}]=i\theta$, 
where $\hat{x}$ and $\hat{y}$ are spatial coordinates 
and $\theta$ is a parameter that governs
the noncommutativity of the space which is known as the Moyal plane. 
In finding nontrivial solutions, projection operators are essential. 
The operators of the form $\ket{n}\bra{n}$ are frequently used, where 
$\ket{n}$ is an eigenstate of the number operator $\hN=\hat{a}^{\dagger}\hat{a}$.
When considering a scalar field as a tachyon on a non-BPS D2-brane, it is known that such a noncommutative soliton corresponds to a D0-brane 
\cite{Dasgupta:2000ft, Harvey:2000jt, Harvey:2001yn}.

In \cite{Asakawa:2009yb}, we applied the GMS solitons 
to a $(2+1)$-dimensional gravity on a Moyal plane that has a cosmological constant 
term only, which means that it does not have the Ricci scalar term.
Nevertheless, 
there do exit nontrivial solutions in the theory 
when the noncommutativity between the space coordinates
is imposed. 
In particular, we found a solution whose geometry is a disc 
in a noncommutative Minkowski spacetime with the metric \cite{Asakawa:2009yb}
\be
G_{\mu\nu} = \eta_{\mu\nu}
\big( \ket{0}\bra{0}+\ket{1}\bra{1}+\cdots +\ket{N-1}\bra{N-1} \big).
\label{Minkowski}
\ee
 
Apart from the metric, this spacetime is equivalent to the fuzzy disc,  
which was first referred in \cite{Lizzi:2003ru, Lizzi:2003hz, Lizzi:2006bu}. 
The fuzzy disc is a disc-shaped region in a Moyal plane
and is a fuzzy approximation of a commutative 
disc by matrices with finite degrees of freedom. 
The algebra is a subalgebra of the one characterizing the Moyal plane 
with a $\st$ product and depends on 
the size of the matrices $N$ and the noncommutative 
parameter $\theta$. 
It was introduced for describing the quantum Hall effect as a Chern-Simon theory on that space.
The behavior of the fuzzy disc in various limits
are investigated in \cite{Lizzi:2003ru, Lizzi:2003hz, Lizzi:2006bu}.  
For example, they took the limit of $N\to \infty$ and  $\theta \to 0$ 
with $N\theta$ fixed, which corresponds to recovering a commutative disc
with a radius fixed.

In this respect, the fuzzy disc is a fascinating arena where gravity, condensed matter and D-branes meet together.
However, most of the consideration so far were based on the number operator and projection operators $\ket{n}\bra{n}$, 
which correspond to looking the fuzzy disc by cutting 
it concentrically.
On the other hand, as far as finding nontrivial solutions we mentioned above, 
we can use any kind of projection operators, as is already referred in \cite{Gopakumar:2000zd}. 

Now we explain what we would like to do in this paper. 
The main issue of this paper is to address how a concept of angles 
can be introduced to the fuzzy disc. 
We will show that there is a well-defined notion of the ``angle" operator $\hat{\varphi}$ on the fuzzy disc, 
which is known as the phase operator suggested by Pegg and Barnett
in quantum optics \cite{Pegg1}. 
Together with the number operator $\hN$, we can characterize the fuzzy disc by certain commutation relations.  
The corresponding angle states
which are the eigen states of $\hat{\varphi}$ 
make possible to define another class of projection operators, angular projection operators $\ket{\varphi_m}\bra{\varphi_m}$.
We will use them to construct noncommutative solitons without circular symmetry. 
These angular projection operators are NOT linear combinations of $N$ projection 
operators $\ket{0}\bra{0}$, $\ket{1}\bra{1}$, $\cdots$, $\ket{N-1}\bra{N-1}$, but consist of $\ket{n}\bra{m}$ with 
$n\neq m$. 
Due to this, they pick up fan-shaped regions along a particular discretized direction in the fuzzy disc,
as opposed to $\ket{n}\bra{n}$. 

As far as we know, this is the first usage of ``angles" in the fuzzy disc.
Although we mainly focus on finding new noncommutative soliton solutions of field theories,
it gives a new viewpoint to understand the geometry on the fuzzy disc.
In fact, this point of view connects the degrees of freedom of the 
boundary of a disc to those of its bulk 
as opposed to dividing the disc concentrically by 
the operators with circular symmetry. 
This is why  the concept of angles is thought to 
suggest the idea that the fuzzy disc would be a good tool 
in various fields in physics such as 
the quantum hall effect or black hole microstates, where 
the ``bulk-boundary correspondence" plays a important role.
We will briefly mention these issues and in addition  
a possibility of an experiment 
concerning the fuzzy disc by means of laser physics.


The organization of this paper is as follows. 
In the next section, we review the fuzzy disc 
introduced by the authors of \cite{Lizzi:2003ru, Lizzi:2003hz, Lizzi:2006bu}.  
In Sec.3, we introduce a concept of angles to the fuzzy disc,
by reinterpreting the phase operator of Pegg and Barnett 
\cite{Pegg1, Barnett1, Pegg:1989zz, Vaccaro1}.
Then we define the angular projection operators and study their properties. 
We also refer a fuzzy annulus, which is the fuzzy disc with a hole 
in its center and is one of the variations of the fuzzy disc. 
In Sec.4, we consider several applications.  
First, a scalar field theory on a $(2+1)$-dimensional
spacetime with noncommutative space coordinates is considered. 
The large-noncommutativity limit of the 
equation of motion for that theory is derived and 
new fan-shaped soliton solutions are shown. 
Second, we consider a gravitational theory on the noncommutative spacetime
that we gave in \cite{Asakawa:2009yb}. 
The last section is for discussions and some implications 
on possible applications of the fuzzy disc to black hole micro states 
and a realization of the analogues of fuzzy objects 
using the Gaussian beam in laser physics.


\section{Review of the Fuzzy Disc}

The fuzzy disc was first introduced in \cite{Lizzi:2003ru, 
Lizzi:2003hz, Lizzi:2006bu}, which is a disc-shaped region 
in a two-dimensional Moyal plane.
This is also a fuzzy approximation of the ordinary (i.e., commutative) 
two-dimensional disc by replacing functions on it with 
finite $N\times N$ matrices.

\subsection{Moyal plane}

Let us start with a Moyal plane,
which is a flat space with noncommutative coordinates 
satisfying the Heisenberg commutation relation\footnote{
The parameter of noncommutativity $\theta$ 
we use through this paper is twice as large as the one 
used in \cite{Lizzi:2003ru}.},
\be
[\hat{x}, \hat{y}]=i\theta.
\ee
The algebra of functions on this noncommutative plane is an operator algebra 
$\hat{\cal A}$ generated by
$\hat{x}$ and $\hat{y}$, acting on a Hilbert space
${\cal H}=l^2={\rm span}\{\ket{0},\ket{1},\cdots\}$.
Here as in standard quantum mechanics, $\ket{n}$ is an eigenstate of 
the number operator 
\be
\hN \ket{n} = n\ket{n}, \quad \hN \equiv \daga \hat{a},
\ee
where the creation and annihilation operators are defined as 
\be
\hat{a} = \frac{\hat{x}+i\hat{y}}{\sqrt{2\theta}}, \quad
\hat{a}^{\dagger} = \frac{\hat{x}-i\hat{y}}{\sqrt{2\theta}}.
\label{HO}
\ee
Then, any operator is expressed by the matrix elements in this basis as 
\be
\hO = \sum_{m,n=0}^{\infty} O_{mn} \ket{m}\bra{n},
\ee
where $O_{mn}$ is a  $c$-number.

Instead of working with operators, one can also consider functions on the commutative plane with a deformed noncommutative product $\st$ by means of the Weyl-Wigner correspondence.
It associates an operator ${\cal O}_f (\hat{x}, \hat{y})$ 
with a function (symbol) $f(x,y)$ such that 
the product of operators is equivalent to the deformed product as 
${\cal O}_f {\cal O}_g ={\cal O}_{f\st g}$.
Note that there is an ambiguity in this correspondence due to operator ordering.
This implies an ambiguity in defining a deformed product.
For instance, operators with the Weyl ordering  
used in \cite{Gopakumar:2000zd} 
are mapped to functions with the Moyal product,
while operators with the normal ordering are mapped 
to functions with the Wick-Voros product.
Both of the orderings are equivalent \cite{Galluccio:2008wk}. 
We adopt the normal ordering through this paper following \cite{Lizzi:2003ru, Lizzi:2003hz, Lizzi:2006bu} %
\footnote{
The symbol used in \cite{Gopakumar:2000zd} is known as the Laguerre-Gaussian function in laser physics. 
We will comment on this issue in the discussion. 
}. 

To be more precise, let us consider a normal-ordered operator $\hat{f}$ 
that is expanded in terms of the creation and annihilation 
operators as 
\be
\hat{f} \equiv \sum_{m,n=0}^{\infty} f_{mn}^{\mbox{\tiny Tay}}
\hat{a}^{\dagger m} \hat{a}^n.
\ee
The symbol map based on the Weyl-Wigner correspondence 
associates $\hat{f}$ with a function $f$ as
\be
f(z, \bz) = \bra{z}\hat{f}\ket{z},
\ee
where $\ket{z}$ is a coherent state satisfying 
$\hat{a}\ket{z} = (z/\sqrt{2\theta})\ket{z}$.
Then, a product of two operators $\hat{f}\,\hat{g}$ is expressed by the Wick-Voros product 
\be
(f\st g) (z,\bz) =e^{2\theta \partial_{\bz'} \partial_{z''}}f(z',\bz')g(z'',\bz'')|_{z=z'=z''}. 
\ee
As an example, let us consider a set of orthogonal projection operators
\be
\hat{p}_n= \ket{n}\bra{n} \quad (n=0, 1, \cdots), 
\ee
which satisfy $\hat{p}_m^\dagger=\hat{p}_m$ and 
\be
\hat{p}_m \, \hat{p}_n = \delta_{mn}\,  \hat{p}_n.
\ee
The corresponding function to the projection operator 
$\hat{p}_n$ is given by
\be
p_n(r) = \bracket{z}{n}\bracket{n}{z}
=e^{-\frac{r^2}{2\theta}}\frac{r^{2n}}{n! (2\theta)^n},
\label{ProjFunc}
\ee
where we used the following equations; 
\ba
\bracket{z}{n}  &=& e^{-\frac{r^2}{4\theta}}\frac{\bz^n}{\sqrt{n! (2\theta)^n}}, \\
\bracket{n}{z}  &=& e^{-\frac{r^2}{4\theta}}\frac{z^n}{\sqrt{n! (2\theta)^n}}.
\ea
We note that the function $p_n(r)$ depends on the radial coordinate 
$r$ only, where 
\begin{align}
z=x+iy=re^{i\varphi}.
\end{align}
Conversely, any circular-symmetric operator $\hat{f}$  
can be expanded 
by $\{\hat{p}_0, \hat{p}_1, \cdots \}$ as 
\be
\hat{f} = \sum_{n=0}^{\infty} f_n \hat{p}_n, 
\ee
where $f_n$ is a  $c$-number. 
Note that the decomposition of the operators 
on a Moyal plane by $\{\hat{p}_0, \hat{p}_1, 
\cdots \}$ is the concentric description of the functions.
Also, since the projection operators satisfies the completeness 
condition 
\be
\sum_{n=0}^{\infty} \hat{p}_n =1, 
\ee
we can recover 
the whole Moyal plane by summing all of the projection 
operators.

\subsection{Fuzzy disc}


The fuzzy disc is defined as a subalgebra of an operator algebra $\hat{\cal A}$ 
on a Moyal plane
 by restricting to $N\times N$ matrices in the number basis.
It is obtained by the projection 
$\hat{\cal A}_N =\hat{P}_N \hat{\cal A} \hat{P}_N$ 
through the rank $N$ projection operator, 
\be
\hat{P}_N= \sum_{n=0}^{N-1}\hat{p}_n =\hat{p}_0 + \cdots + \hat{p}_{N-1}.
\label{completenessofP}
\ee
Any operator in $\hat{\cal A}$ has the form $\hat{P}_N \hat{f}\hat{P}_N$ for $\hat{f}\in \hat{\cal A}$.
Note that this operator $\hat{P}_N$ plays the role of the identity operator $1_N$ in 
$\hat{\cal A}_N$. 
The projection operator also has its corresponding function;
\be
P_N (r) = 
\sum_{n=0}^{N-1} e^{-\frac{r^2}{2\theta}}\frac{r^{2n}}{n! (2\theta)^n}
=\frac{\Gamma (N,r^2/(2\theta))}{\Gamma (N)},
\label{incomplete gamma}
\ee  
where $\Gamma (n,x)$ is the incomplete gamma function.
This function is roughly a radial step function that picks up a disc-shaped region around the origin $z=0$ with radius $R=\sqrt{2N\theta}$.
\begin{figure}[tb]
  \begin{center}
   \includegraphics[scale=0.45, bb=0 0 259 214]{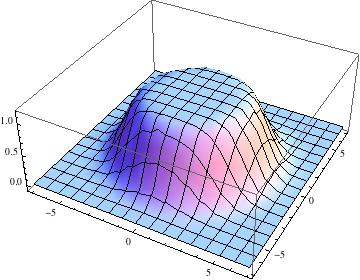}
   \hspace{3cm}
    \includegraphics[scale=0.45, bb=0 0 259 214]{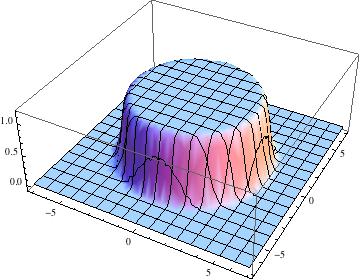}
    \caption{The fuzzy disc for $N=10, \theta =1$ (left) and  
    $N=100, \theta =0.1$ (right).} 
    \label{FuzzyDisc1}
  \end{center}
\end{figure}
This is why the authors of \cite{Lizzi:2003ru} called $\hat{\cal A}_N$ as the fuzzy disc.
As examples of the fuzzy discs, those with $N=10, \ \theta =1$  and with
$N=100, \ \theta =0.1$ are shown in Figure \ref{FuzzyDisc1}. 

However, as emphasized in \cite{Lizzi:2003ru}, $\hat{\cal A}_N$ is isomorphic to the matrix algebra and any fuzzy space is isomorphic to it so that it is not apparent that the space is actually disc-shaped when one works with the matrix algebra.
To overcome this difficulty, the authors of \cite{Lizzi:2003ru} investigated the behavior of 
the fuzzy disc by taking various limits, 
and claimed that the disc should be recognized in the correlated limit of $\theta$ and $N$ keeping the radius $R$ fixed. 

\begin{figure}[b]
  \begin{center}
   \includegraphics[scale=0.55, bb=0 0 259 214]{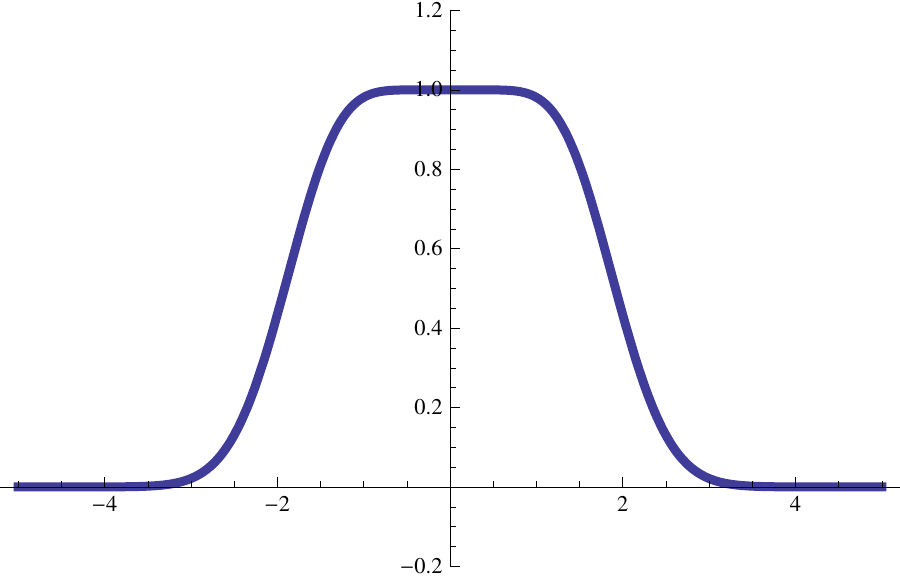}
   \hspace{3cm}
    \includegraphics[scale=0.55, bb=0 0 259 214]{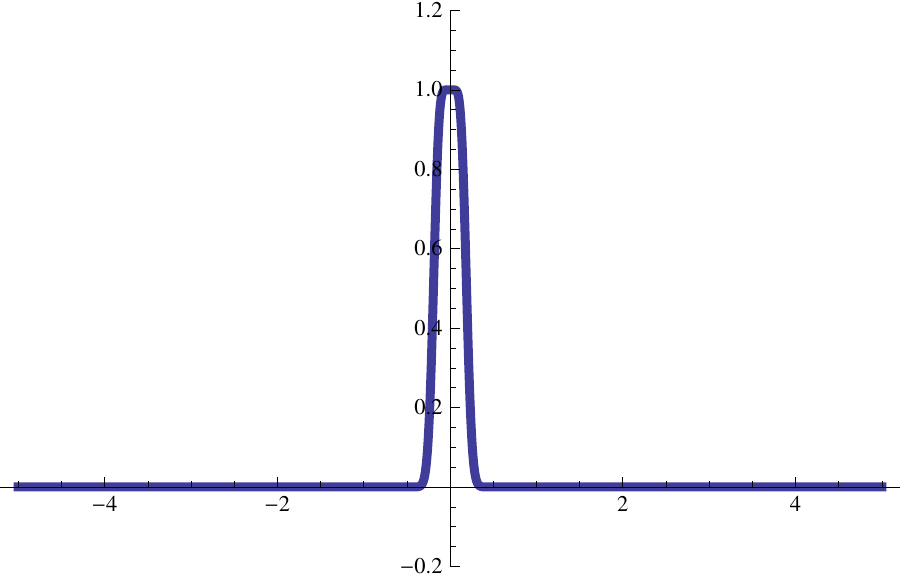}
    \caption{The section of the fuzzy disc for $N=4, \theta =1$ (left) and  
   for $N=4, \theta =0.01$ (right).} 
    \label{FuzzyDisc2}
  \end{center}
\end{figure}

First, they considered the commutative limit; $\theta \to 0$ with 
finite $N$ fixed. 
As the radius of the fuzzy disc is given by $R=\sqrt{2N\theta}$, 
this limit makes the fuzzy disc to a one point in the two-dimensional 
space, which is shown in Figure \ref{FuzzyDisc2}. 

\begin{figure}[tb]
  \begin{center}
   \includegraphics[scale=0.55, bb=0 0 259 214]{2dimFD_N4Theta1.pdf}
   \hspace{3cm}
    \includegraphics[scale=0.55, bb=0 0 259 214]{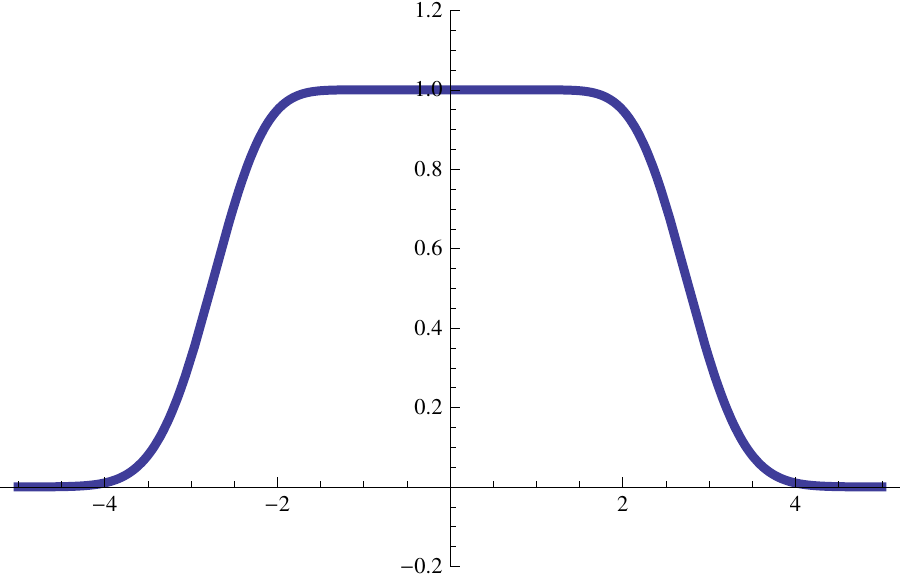}
    \caption{The section of the fuzzy disc for $N=4, \theta=1$ (left) 
    and  
    $N=10, \theta=1$ (right).} 
    \label{FuzzyDisc3}
  \end{center}
\end{figure}

Second, they took $N\to \infty$ with finite $\theta$ fixed. 
As $\theta$ is not zero, the space remains noncommutative. 
Then we see the whole Moyal plane is reproduced, 
which is shown in Figure \ref{FuzzyDisc3}.

\begin{figure}[b]
  \begin{center}
   \includegraphics[scale=0.6, bb=0 0 259 214]{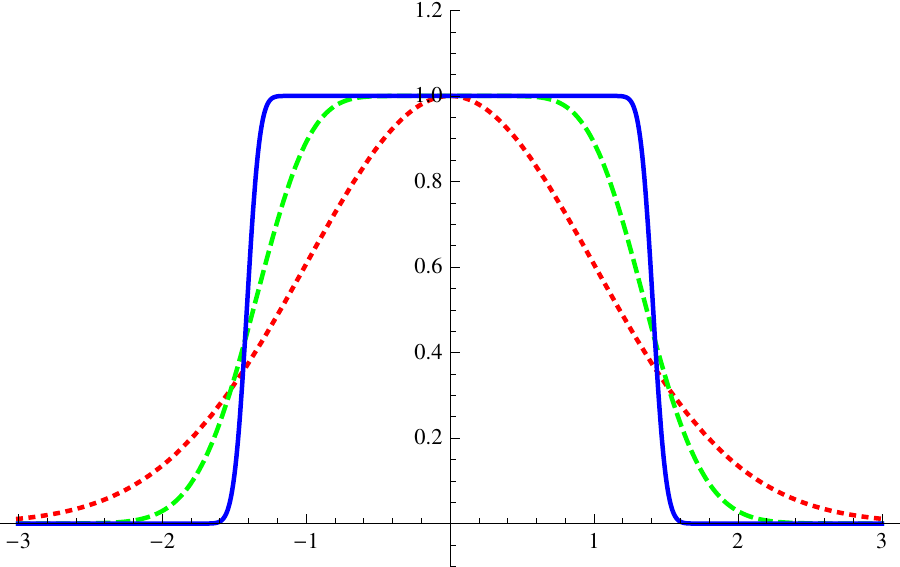}
    \caption{The section of the fuzzy disc for $N=1, \theta=1$ (red, dotted), 
    $N=10, \theta=0.1$ (green, dashed) and  
     $N=100, \theta =0.01$ (blue, solid).} 
    \label{FuzzyDisc4}
  \end{center}
\end{figure}

One more limit they investigated is that $N \to \infty$ and $\theta \to 0$ 
with $N\theta$ fixed. This corresponds to the situation where 
the radius of the disc does not change and the noncommutativity 
disappears. So we obtain the disc with the finite radius 
and commutative 
space coordinates. This feature appears in the slope of the fuzzy disc. 
As shown in Figure \ref{FuzzyDisc4}, 
the slope of the boundary of the fuzzy disc becomes steeper  
and steeper as $N\to \infty$ and $\theta \to 0$ as explained 
in \cite{Lizzi:2003ru}. 
The disc becomes the flat and finite region with radius $R=\sqrt{2N\theta}$.  

They also referred to the edge state in the same paper. 
The edge state is a localized state of a quantum system with boundary. 
The most famous example of that would be the edge state 
in the quantum Hall effect. 
As is well known, the Hall conductivity is quantized so exactly
that it is thought to have the topological origin of the boundary 
of the system. 
In particular cases, the behaviors of the boundary determine
the whole systems and this is known as 
the bulk-boundary correspondence \cite{Hatsugai1, Hatsugai2}.
The fuzzy disc has this edge state as one of its states 
and the relation between the edge states 
and the noncommutativity is discussed.

\section{Angles in the Fuzzy Disc and Angular Projection Operators}

In this section, we introduce a concept of angles to the fuzzy disc.
After briefly reviewing the phase operator developed in quantum optics, 
we show that this operator can be regarded 
as the appropriate angle operator.
Then, we focus on new angular projection operators that 
corresponds to fan-shaped regions.

\subsection{Pegg-Barnett's Phase Operator}

Here we would like to give a short review of the phase operator 
suggested by Pegg and Barnett, following \cite{Matsuoka}.

In the quantum theory of electromagnetic fields, it is known and widely used 
that there is an uncertainty relation $\Delta N \Delta \varphi \ge 1/2$ 
for the photon number and the phase. 
Historically, Dirac first referred such a relation, by assuming that there is a Hermitian operator $\hat{\varphi}$ which satisfies a canonical commutation relation \cite{Dirac:1927},
\be
[\hN, \hat{\varphi}] = i, 
\ee
where $\hN$ is the photon-number operator and $\hat{\varphi}$ is the phase operator
that corresponds to a $c$-number $\varphi$ which appears in the classical radiation field
$a = \sqrt{N}e^{i\varphi}$ \cite{Susskind:1964zz, Loudon}. 
If there  would be such a Hermitian operator $\hat{\varphi}$, a unitary operator 
$\exp( i \hat{\varphi})$ also would exist, and vice versa.
However, Susskind and Glogower showed that there is not such a unitary operator,  
therefore the Hermitian operator corresponding to $\varphi$ can not exist, 
either \cite{Susskind:1964zz}. 

To see the essence of this argument, 
let us consider the polar decomposition of the photon annihilation operator
\be
\hat{a} = (\hN+1)^{1/2} \hat{S}^\dagger,
\label{aandU}
\ee
where $\hat{S}$ is the shift operator defined by $\hat{S}\ket{n}=\ket{n+1}$ for all $n$.
On the other hand, its Hermitian conjugate satisfies
\ba
\hat{S}^\dagger \ket{n} 
&=& 
\left\{ \begin{array}{cc}
\ket{n-1}  & (n\ge 1), \\
 0 & (n=0).  
 \end{array} \right.
\ea
Clearly it leads $\hat{S}^{\dagger}\hat{S}=1$ but $\hat{S}\hat{S}^{\dagger} =1-\ket{0}\bra{0} \neq 1$. 
Thus, $\hat{S}$ can not be unitary (such an operator is called an isometry in operator algebraic language).
This means that there is neither the Hermitian 
operator $\hat{\varphi}$ 
nor the unitary operator of the form $\hat{S} =\exp(-i\hat{\varphi})$. 
Along this line, the construction of a phase operator had been thought 
to be impossible, but Pegg and Barnett changed this situation 
\cite{Pegg1, Barnett1, Pegg:1989zz}. 
Their idea is based on a phase state $\ket{\varphi}$ 
that Loudon suggested in \cite{Loudon},
\be
\ket{\varphi} = \lim_{N\to \infty} \frac{1}{\sqrt{N}} \sum_{n=0}^{N-1}
e^{in\varphi} \ket{n}. 
\ee 

The fact they found is that there is a well-defined phase operator if the Hilbert space is restricted to its finite dimensional subspace.
Let ${\cal H}_N={\rm span}\{\ket{0},\cdots,\ket{N-1}\}$ be such a subspace. The construction of them is as follows. 
They first defined a phase state whose eigenvalue is $\varphi_0$ as 
\be
\ket{\varphi_0} = \frac{1}{\sqrt{N}} \sum_{n=0}^{N-1}
e^{in\varphi_0} \ket{n}. 
\ee 
Then it is easy to find the other $(N-1)$ states $\ket{\varphi_m}$ that satisfy 
$\bracket{\varphi_m}{\varphi_n}=\delta_{mn}$ as
\be
\ket{\varphi_m} =  \frac{1}{\sqrt{N}} \sum_{n=0}^{N-1}
e^{in\varphi_m} \ket{n}, 
\ee 
where 
\be
\varphi_m = \varphi_0 + \frac{2\pi}{N}m \quad (m=0, 1, \cdots, N-1). 
\ee
Inversely, the expansion of $\ket{n}$ by $\ket{\varphi_m}$ is 
given by 
\be
\ket{n} = \frac{1}{\sqrt{N}} \sum_{m=0}^{N-1}e^{-in\varphi_m} \ket{\varphi_m}. 
\ee
Thereby a set of $N$ states 
$\{\ket{\varphi_0}, \ket{\varphi_1}, \cdots, \ket{\varphi_{N-1}}\}$ 
forms an orthonormal basis for the $N$-dimensional subspace ${\cal H}_N$.
By using them, one can define an operator of the following form
\be
\hat{\varphi} = \sum_{m=0}^{N-1} 
\varphi_m\ket{\varphi_m}\bra{\varphi_m}. 
\ee
We can call this operator $\hat{\varphi}$ the phase operator 
that has $\ket{\varphi_m}$ as its eigen state and 
$\varphi_m$ as the corresponding eigen value. 
Also, it is a Hermitian operator acting on ${\cal H}_N$.
It can be expanded by the number states as 
\be
\hat{\varphi} 
= \left(\varphi_0 + \frac{(N-1)\pi}{N}\right) 1_N + \frac{2\pi}{N}
\sum_{n\neq n'}\frac{e^{i(n'-n)\varphi_0}}{e^{2\pi i(n'-n)/N}-1}
\ket{n'}\bra{n}. 
\ee
Now we obtain the unitary operator based on $\hat{\varphi}$ as 
\be
\hU = \exp({i\hat{\varphi}}), 
\ee
where its eigenstates are $\ket{\varphi_m}$ 
and the corresponding eigenvalues are 
$e^{i \varphi_m}$ for $m=0,1,\cdots, {N-1}$.
When acting on the number basis $\ket{n}$ ($n=1, 2,  \cdots, N-1$), 
this operator behaves as the (inverse) shift operator, while we have 
$\hU\ket{0}=e^{iN\varphi_0}\ket{N-1}$.
That is, $\hU$ is a cyclic operator, 
\be
\hU = \ket{0}\bra{1}+\ket{1}\bra{2} + \cdots + \ket{N-2}\bra{N-1}
+e^{iN\varphi_0}\ket{N-1}\bra{0}.
\ee
In this sense, the singularity for the shift operator $\hat{S}$ at $n=0$ 
which prevents from the construction of the Hermitian operator 
is resolved. 

In the Pegg-Barnett formalism, all quantities are initially defined and calculated on the finite dimensional Hilbert space ${\cal H}_N$, and then the $N\to \infty $ limit has to be taken. 
The justification of this formalism has been argued and tested experimentally, 
which leads to physically admissible results for various quantum states of the radiation field so far
\cite{Abe1995, Atakishiyev:2010zz, Fujikawa:1995hh, Fujikawa:1994yf}.



\subsection{Angles in the fuzzy disc}

The construction of the phase states and the phase operator in the previous section 
can be straightforwardly applicable to our setting on the fuzzy disc.
Unlike quantum optics, we regard the phase 
$\varphi_m$ as an angle 
in a Moyal plane. 
From now on,  we also call 
the state $\ket{\varphi_m}$ 
the angle state rather than the phase state. 
Thus we define the angle states in the Hilbert space ${\cal H}_N=\hat{P}_N{\cal H}={\rm span}\{\ket{0},\cdots,\ket{N-1}\}$ of the fuzzy disc as 
\be
\ket{\varphi_m} =  \frac{1}{\sqrt{N}} \sum_{n=0}^{N-1}
e^{in\varphi_m} \ket{n}, 
\ee 
and the angle operator as
\be
\hat{\varphi} =  \sum_{m=0}^{N-1} \varphi_m \ket{\varphi_m}\bra{\varphi_m}.
\ee 
Here the eigenvalues 
\be
\varphi_m =\frac{2\pi}{N}m \quad (m=0, 1, \cdots, N-1),
\ee
are $N$ discrete angles that are uniformly distributed in the interval $[0,2\pi]$,
and evidently periodic $\varphi_{m+N}=\varphi_m$.
They are thus a fuzzy approximation of the continuous angles in a commutative disc.
We have set $\varphi_0=0$ as compared to the previous subsection for simplicity.
Note also that there is no problem concerning the $N\to \infty $ limit at this stage, because the fuzzy disc is defined for a finite $N$, as opposed to quantum optics.

Next let us define the number operator restricted to 
${\cal H}_N={\rm span}\{\ket{0},\cdots,\ket{N-1}\}$ as
\be
\hat{N} =  \sum_{n=0}^{N-1}n \ket{n}\bra{n}, 
\ee 
where we have used 
the same symbol as in the whole Hilbert space ${\cal H}=l^2$.
The eigen value $n$ works as the radial coordinate on the fuzzy disc
(more precisely, the radial coordinate operator should be defined by 
$\hat{r}=\sqrt{2\theta\hat{N}}$.).

Since the original Moyal plane is noncommutative, these two coordinate operators 
$\hat{N}$ and $\hat{\varphi}$ are noncommuting with each other, 
but the form of their commutation relation is rather complicated.
In order to express the noncommutativity, it is more instructive to introduce two more operators defined by 
\begin{align}
\hat{V}&:=e^{i\frac{2\pi}{N}\hat{N}}=\sum_{n=0}^{N-1}e^{i\frac{2\pi}{N}n}\ket{n}\bra{n}
=\sum_{m=0}^{N-1}\ket{\varphi_{m+1}}\bra{\varphi_m},\label{V}\\
\hat{U}&:=e^{i\hat{\varphi}}=\sum_{m=0}^{N-1}e^{i\varphi_m}\ket{\varphi_m}\bra{\varphi_m}
=\sum_{n=0}^{N-1}\ket{n-1}\bra{n},\label{U}
\end{align}
where the latter $\hat{U}$ has already been given in the previous subsection.
To prove each equality in these equations, the relation between two orthonormal bases 
\be
\bracket{n}{\varphi_m}=\frac{1}{\sqrt{N}}e^{in\varphi_m}
\ee
in ${\cal H}_N$, and an identity 
\begin{align}
\sum_{n=0}^{N-1}e^{i\frac{2\pi}{N}kn}=N\delta_{k,0}. 
\end{align}
for all $k \in \{0,\cdots,N-1\}$ are frequently used%
\footnote{
In the last expression in (\ref{U}), the $n=0$ term should be understood as $\ket{-1}:=\ket{N-1}$. 
But it is just for notational simplicity, and this does not mean any periodicity in the number basis as opposed to the angle basis.
}.

These operators $\hat{V}$ and $\hat{U}$ are unitary and 
satisfy  $\hat{V}^N=\hat{U}^N=1_N$.
As seen from the last expression in (\ref{V}), $\hat{V}$ acts on 
the angle states as a unit shift, i.e., the rotation of the disc, while $\hat{U}$ behave as a radial shift operator.
It is easy to show the following commutation relations of 
the operators we encountered:
\begin{align}
[\hat{N},\hat{U}]=-\hat{U} + N\hat{U}\hat{p}_0,
\quad [\hat{\varphi},\hat{V}]=\frac{2\pi}{N}\hat{V},
\quad \hat{U}\hat{V}=e^{\frac{2\pi i}{N}}\hat{V}\hat{U}.
\label{commutation relations}
\end{align}

We would like to characterize the fuzzy disc in terms of these operators.
Before this, recall that the third equation in (\ref{commutation relations}) 
is nothing but  the defining commutation relation of the fuzzy torus.
Any operator of the form $\hat{f}=\sum_{m,n=0}^{N-1} f_{mn} \hat{U}^m \hat{V}^n$, generated by two unitaries $\hat{U}$ and $\hat{V}$, can be regarded as a function on that torus, while $\hat{N}$ and 
$\hat{\varphi}$ play the role of derivatives, as seen from first two commutation relations.
What makes this algebra to be that of a torus should again be considered together with the limiting 
procedure $N\to\infty$. 

On the contrary, we will now argue that 
the algebra for the fuzzy disc is generated by a operator
\be
\hat{z}=\hat{U}\sqrt{2\theta \hat{N}}=e^{i\hat{\varphi}}\hat{r}, 
\label{operator z}
\ee
and its hermitian conjugate $\hat{z}^\dagger$, 
subject to the commutation relation
\be
[\hat{z}, \hat{z}^\dagger] =2\theta(1-N\hat{p}_{N-1}),
\label{com rel}
\ee
and any operator corresponding to a function on a fuzzy disc has the form
\begin{align}
\hat f =\sum_{k,l=0}^{N-1} f_{kl} \hat{z}^{\dagger k} \hat{z}^l.
\label{fuzzy disc element}
\end{align}
Before showing this, we would like to give a few remarks are in order.
First, (\ref{operator z}) is also seen as the polar decomposition of 
the generator $\hat{z}$, 
written by the Hermitian $\hat{N}$ and the unitary $\hat{U}$ operator, as opposed to 
the creation operator $\hat{a}$.
Note that $\hat{z}^\dagger \hat{z}=2\theta \hat{N}$.
Next, the $\hat{p}_{N-1}$ term in (\ref{com rel}) 
guarantees that the commutator is traceless ${\rm Tr} [\hat{z}, \hat{z}^\dagger] =0$.
Note also $\hat{U}\hat{p}_0=\hat{p}_{N-1}\hat{U}$.

To show the statement above, recall that 
the definition  of the 
fuzzy disc algebra 
is $\hat{\cal A}_N =\hat{P}_N \hat{\cal A} \hat{P}_N$. 
Thus, by the Weyl-Wigner correspondence, 
any function $f(z,\bar{z})$ on a plane gives an operator in $\hat{\cal A}_N$ of the form 
\begin{align}
\hat f =\hat{P}_N \left(\sum_{k,l =0}^\infty f_{k,l} \hat{a}^{\dagger k} \hat{a}^l \right) \hat{P}_N. 
\end{align}
We show that it reduces to (\ref{fuzzy disc element}).
First, a function $f=z$ corresponds to 
\begin{align}
\sqrt{2\theta} \hat{P}_N \hat{a} \hat{P}_N 
=\sum_{n=1}^{N-1} \sqrt{2\theta n} \ket{n-1}\bra{n}
=\hat{U}\sqrt{2\theta \hat{N}}=\hat{z}.
\end{align}
Similarly, for $f=z^l$ ($0\le l \le N-1$), we obtain an operator
\begin{align}
(2\theta)^{\frac{l}{2}} \hat{P}_N \hat{a}^l \hat{P}_N 
=\left(\sqrt{2\theta}\hat{P}_N \hat{a} \hat{P}_N\right)^l
=\hat{z}^l,
\end{align}
where in the first equality, $1_N=\hat{P}_N +(1-\hat{P}_N) $ are inserted,
and the identity $(1-\hat{P}_N)\hat{a} \hat{P}_N=0$ was used.
It is also shown that $\hat{P}_N \hat{a}^l \hat{P}_N =0$ for $N \le l$.
In the same way, we obtain $\hat{z}^{\dagger k}$ for $f=\bar{z}^k$, 
and $\hat{z}^{\dagger k} \hat{z}^l$ for $f=\bar{z}^k z^l$
($0\le k, l \le N-1$), 
where another identity $\hat{P}_N \hat{a}^\dagger (1-\hat{P}_N)=0$ is also used in these cases.
This shows the validity of (\ref{fuzzy disc element}).
It is regarded as a variant of the 
Weyl map from functions on the plane to 
the fuzzy disc algebra, but it is of course 
not one to one because 
higher frequency modes than $N-1$ such as $z^N$ are projected out.

Next, let us consider the symbol map of operators (\ref{fuzzy disc element}), 
which are functions written by $(z,\bar{z})$ or $(r,\varphi)$ on the whole plane, 
where $z=re^{i\varphi}$.
As is easily shown, the symbol of the generator $\hat{z}$ is 
\begin{align}
z(r,\varphi) &=\bra{z}\hat{z}\ket{z}\nonumber\\
&=\sum_{n=1}^{N-1} \sqrt{2\theta n} \bracket{z}{n-1}\bracket{n}{z}\nonumber\\
&=\sum_{n=1}^{N-1} e^{-\frac{r^2}{2\theta}}\frac{r^{2(n-1)}}{(n-1)! (2\theta)^{n-1}}z \nonumber\\
&=P_{N-1}(r)z.
\end{align}
This shows that the symbol behave as the original function $z$, but weighted by 
the damping factor $P_{N-1}(r)$.
A similar analysis shows that the symbol of $\hat{z}^l$ is $z^l (r,\varphi)=P_{N-l}(r)z^l $,
weighted by a sharper damping factor than $z$.
In general, the symbol of (\ref{fuzzy disc element}) is given by
\begin{align}
f(r,\varphi )&= \sum_{{k,l=0},\,{k\ge l}}^{N-1} f_{kl} P_{N-k}(r) \bar{z}^k z^l 
+\sum_{{k,l=0},\,{k< l}}^{N-1} f_{kl} P_{N-l}(r) \bar{z}^k z^l \nonumber\\
&=\sum_{k,l=0}^{N-1} f_{kl} P_{N-{\rm max}\{k,l\}}(r) \bar{z}^k z^l.
\label{general symbol}
\end{align}
It is the polynomial truncation of the original function $f(z,\bar{z})$ except for 
the weight factors.
All these factors $P_{N-{\rm max}\{k,l\}}(r)$ for finite $k$ and $l$ 
tends to the radial step function in the commutative disc limit $N\to \infty$, $\theta \to 0$ 
with $R^2=2N\theta$ fixed
so that the symbol has actually a value only on the interior of the disc $|z|< R$.
Moreover, in this limit, higher and higher powers of $\hat{z}$ and $\hat{z}^\dagger$ become
allowed operators, and finally the space of symbols tends to that of functions on a disc
$\{f(z,\bar{z})||z|<R\}$ of radius $R$. 
This shows that the Weyl map above is one to one when restricted to functions on the disc.

In this way, the disc variant of the Weyl-Wigner correspondence is obtained in terms of 
operators $\hat{z}$ and $\hat{z}^\dagger$, or equivalently $\hat{r}$ and $\hat{\varphi}$.
It is now clear that the commutation relation (\ref{com rel}) 
corresponding to the relation $[z,\bar{z}]_\st =2\theta$ for the Wick-Voros product. 
The use of these operators would be useful to analyze the structure of the fuzzy discs further,
but we do not proceed in this paper and we focus on angular projection operators which will be introduced in the next subsection.

\subsection{Angular projection operators}
\begin{figure}[tb]
  \begin{center}
   \includegraphics[scale=0.38, bb=0 0 259 214]{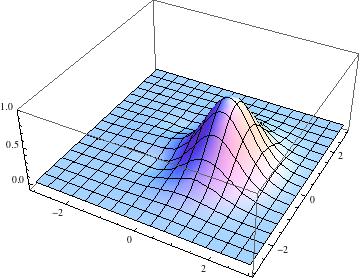}
   \hspace{3cm}
    \includegraphics[scale=0.38, bb=0 0 259 214]{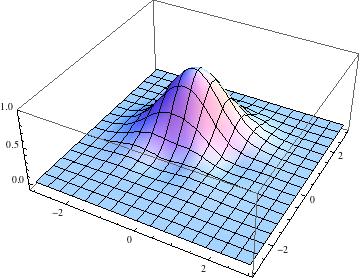}
    \caption{The functions $\pi^{(3)}_0$ (left) 
    and $\pi^{(3)}_1$ (right) for $N=3$.} 
    \label{N03m0and1}
  \vspace{1.5cm}
   \includegraphics[scale=0.38, bb=0 0 259 214]{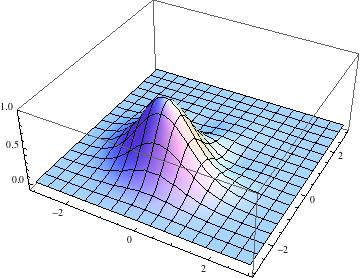}
   \hspace{3cm}
    \includegraphics[scale=0.38, bb=0 0 259 214]{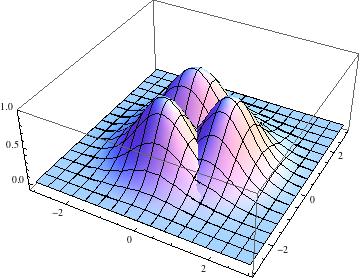}
    \caption{The function $\pi_2^{(3)}$ (left) 
    and all functions $\pi_k^{(3)}$'s (right).}
    \label{N03m2andAll}
  \end{center}
\end{figure}

In the Hilbert space ${\cal H}_N$, let us define an angular projection operator 
\be
\hat{\pi}_k := \ket{\varphi_k}\bra{\varphi_k} 
=\frac{1}{N} \sum_{m,n=0}^{N-1}
e^{i(m-n)\varphi_k} \ket{m}\bra{n},
\label{AngularProj}
\ee
that picks up a particular eigenstate $\ket{\varphi_k}$ of the angle operator.
Because of the orthonormality of the angle states, these projections are 
orthogonal each other:
\be
\hat{\pi}_k \hat{\pi}_l=\delta_{kl}\hat{\pi}_l.
\ee
We also denote it as $\hat{\pi}^{(N)}_k$ when we would like to emphasize that it is an operator 
acting on the $N$-dimensional Hilbert space ${\cal H}_N$.   
By the Weyl-Wigner correspondence, the corresponding function 
$\pi_k(r,\varphi)$
on a plane to the angular projection operator
$\hat{\pi}_k$
is obtained as
\be
\pi_k (r, \varphi) = \frac{1}{N}\sum_{m,n=0}^{N-1}
e^{-\frac{r^2}{2\theta}}
\frac{r^{m+n} }{\sqrt{m!n!(2\theta)^{m+n}}}
e^{-i(m-n)(\varphi-\varphi_k)},
\label{AngularProjFunc}
\ee
where we used the polar coordinates $z=re^{i\varphi}$. 
It is a real function in accordance with Hermiticity. 

As for the projection operators $\hat{p}_n$, the sum of 
all angular projection operators $\hat{\pi}_k$ for a given $N$ satisfies the completeness relation in the Hilbert space 
${\cal H}_N$.
That is,  
\be
\hat{P}_N= 1_N=\hat{\pi}_0 +\cdots +\hat{\pi}_{N-1}.
\label{completeness1}
\ee
By the Weyl-Wigner correspondence, it also maps to a function, which is the radial ``step function" with radius $R=\sqrt{2N\theta}$ on the plane given in (\ref{incomplete gamma}).

\begin{figure}[t]
  \begin{center}
   \includegraphics[scale=0.38, bb=0 0 259 214]{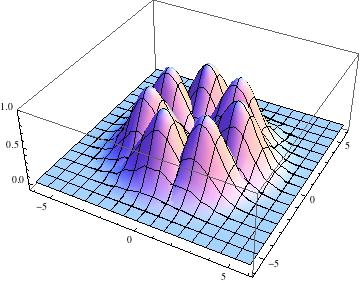}
   \hspace{2.5cm}
    \includegraphics[scale=0.38, bb=0 0 259 214]{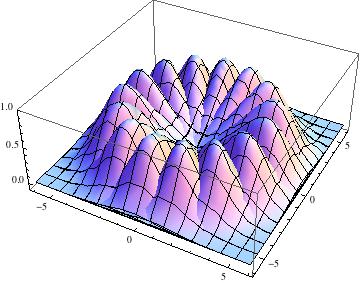}
    \caption{All functions $\pi_k^{(N)}$'s for $N=7$ (left) and for $N=15$ (right). 
    We set $\theta =1$ here. }
     \label{N06AllandN15All}
  \vspace{3cm}
   \includegraphics[scale=0.35, bb=0 0 259 214]{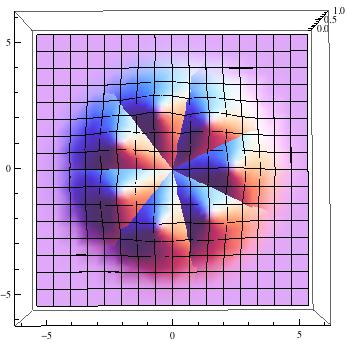}
   \hspace{3cm}
    \includegraphics[scale=0.35, bb=0 0 259 214]{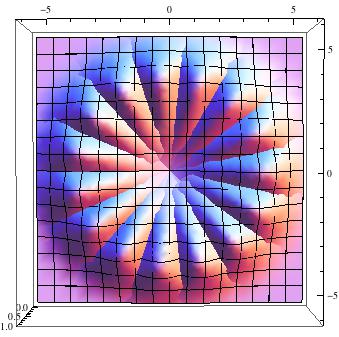}
    \caption{The top views of all $\hPsi_k^{(N)}$'s for $N=7$ (left) 
    and for $N=15$ (right) .    We set $\theta =1$ here. }
     \label{N06AllandN15AllAbove}
  \end{center}
\end{figure}

The fuzzy disc can be considered as a set of $N$ ``points" 
with each point having a unit area $2\pi\theta$.
The two completeness relations in 
(\ref{completenessofP}) and (\ref{completeness1}) correspond to two different decompositions of the same fuzzy disc.
In the former case, a ``point" has roughly a shape of annulus distinguished by its radius, 
while in the latter case, a point is distinguished by its angle. 
Here we show how ``points" are distributed on the plane by the Weyl-Wigner correspondence in Figure \ref{N03m0and1} - 
Figure \ref{N06AllandN15AllAbove}. 

In Figure \ref{N03m0and1} and \ref{N03m2andAll} we show 
the profiles of three functions 
$\pi_0^{(3)}$, $\pi_1^{(3)}$ and $\pi_2^{(3)}$ 
for $N=3$. 
In the right figure of Figure \ref{N03m2andAll}, we draw these three functions simultaneously (not the sum of them), i.e., 
the feet of them overlap each other, therefore, 
at each point $(r,\varphi)$ only the maximal value among 
them is visible. 
This shows that there are three peaks, which are separated 
by $2\pi/3$ each other in the angular direction. 
This behavior is valid for an arbitrary $N$, that is, 
the disc is divided into $N$ fan-shaped regions according to $N$ functions,
in which each function is peaked at the angle $\varphi =\varphi_k$.
In Figure \ref{N06AllandN15All} and \ref{N06AllandN15AllAbove},
the case of $N=7$ and $15$ are shown. 


One can also repeat the analysis in \S 2 in considering three limits, but in the way of distinguishing each point. 
\begin{enumerate}
\item $\theta\to 0$ with $N$ fixed: the point limit\\
This limit is the small radius limit.
As shown in Figure \ref{AngNC1}, each fan-shaped ``point" shrinks to the origin but there are still  distinguished $N$ points.
\begin{figure}[t]
  \begin{center}
   \includegraphics[scale=0.35, bb=0 0 259 214]{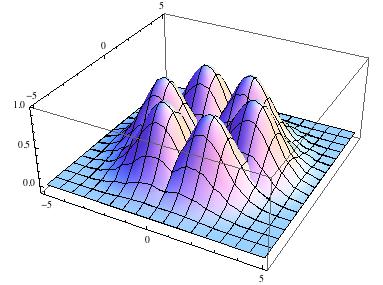}
   \hspace{3cm}
    \includegraphics[scale=0.35, bb=0 0 259 214]{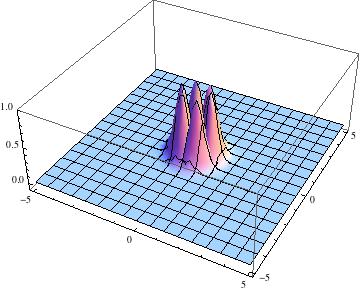}
    \caption{All of the six functions for $N=6, \theta =1$ (left) 
    and for $N=6, \theta =0.1$ (right).} 
    \label{AngNC1}
  \vspace{2cm}
   \includegraphics[scale=0.35, bb=0 0 259 214]{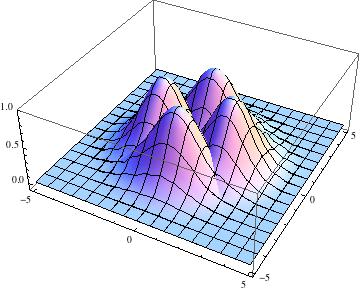}
   \hspace{2cm}
    \includegraphics[scale=0.35, bb=0 0 259 214]{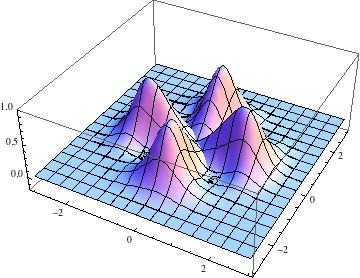}
    \hspace{2cm}
    \includegraphics[scale=0.35, bb=0 0 259 214]{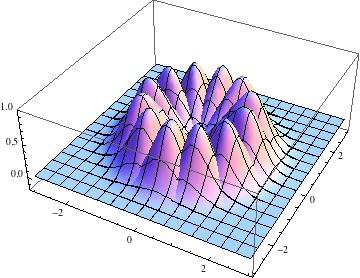}
    \caption{All of the four functions for $N=4, \theta =1$ (left) 
    and four of the twelve functions for $N=12, \theta =1$ (center) 
    whose peaks are in the same positions as in the left figure. 
	The right figure shows all of the twelve functions for $N=12, \theta =1/3$.} 
        \label{AngNC2}
  \end{center}
\end{figure}

\item $N\to \infty$ with $\theta$ fixed: the noncommutative plane limit\\
This limit corresponds to the large radius limit with increasing the degrees of freedom $N$
as we have seen in Figure \ref{N06AllandN15All} where the $N=7$ and $N=15$ 
cases were given. 
However, because of the problem of the phase operator at $N\to \infty$, whether this limit recovers the full Moyal plane should be justified by more careful analysis. 
\item $N\to\infty$ and $\theta \to 0$ with $N\theta$ fixed: 
the commutative disc limit\\
In this limit, the radius $R=\sqrt{2N\theta}$ of the disc is fixed but the noncommutativity disappears. 
Thus we will obtain the commutative disc.
As shown in Figure \ref{AngNC2}, the number of points increases as $N$ grows, and the area shared by each point decreases.

\end{enumerate}
We conclude this section by a remark.
As we have already stated, the fuzzy disc is a collection of $N$ points,
and a point given by an angular projection operator corresponds to a fan-shaped region, 
just like cutting a cake into $N$ pieces.
This is contrasted to a point given by a radial projection operator, 
whose form is similar to a baum-kuchen.
Our cutting would be useful to applying the fuzzy disc to some physical models, such as 
the quantization of physical quantities on the boundary, e.g., 
the edge states \cite{Hatsugai1, Hatsugai2, Balachandran:2003vm, Pinzul:2001qh} 
or black holes \cite{Digal:2011hf}.
This is because the holography is naturally realized in this picture, i.e., 
the degrees of freedom on the boundary is equal to that of the entire disc. 
This issue is left for a future work. 


\subsection{Fuzzy annulus}

\begin{figure}[t]
  \begin{center}
   \includegraphics[scale=0.45, bb=0 0 259 214]{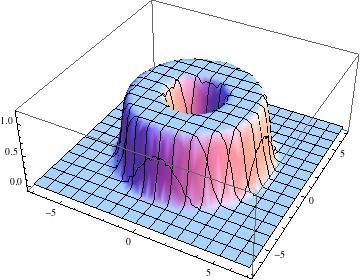}
   \hspace{3cm}
    \includegraphics[scale=0.55, bb=0 0 259 214]{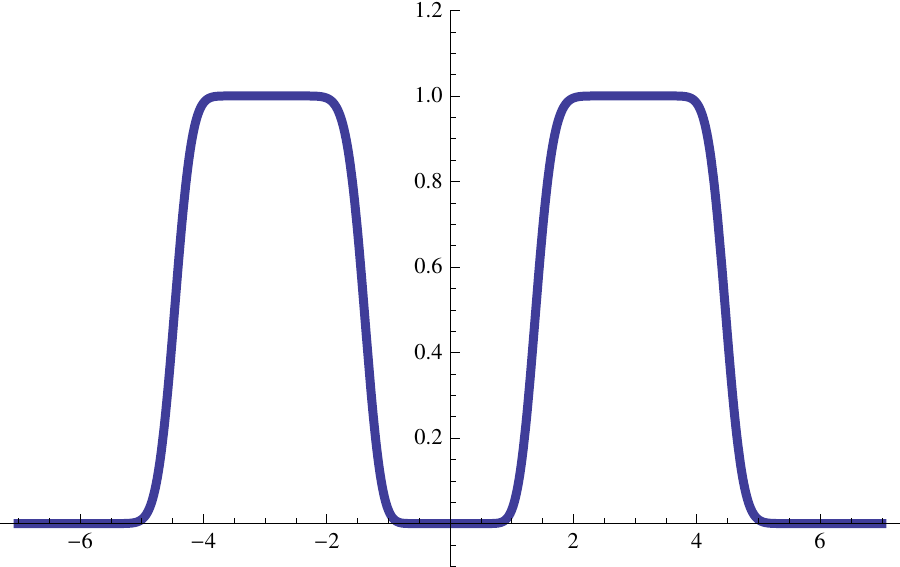}
    \caption{The fuzzy annulus and the cross section of it 
    for $M=100, N=10$ and $\theta =0.1$.}
     \label{FuzzyAnnulus}
  \end{center}
\end{figure}

Here we point out that a fuzzy annulus can be 
constructed in a similar manner as the fuzzy disc.
Let us consider a $N$-dimensional subspace 
${\cal H}^M_N :={\rm span}\{\ket{M},\ket{M+1},\cdots,\ket{M+N-1}\}$ 
of the Hilbert space, starting at $\ket{M}$ for a given $M>0$.
On the whole Hilbert space ${\cal H}=l^2$, it is defined by the projection operator 
\be
\hat{P}_N^M :=\hat{p}_M +\hat{p}_{M+1}+\cdots +\hat{p}_{M+N-1},
\ee
and evidently the subalgebra $\hat{P}_N^M \hat{\cal A} \hat{P}_N^M$ of 
$N\times N$ matrices represents a fuzzy annulus with an inner radius $R_- =\sqrt{2M\theta}$ 
and an outer radius $R_+ =\sqrt{2(N+M)\theta}$.
The corresponding function $P_N^M (r)$ on the plane is shown in 
Figure \ref{FuzzyAnnulus}.
Similar to the fuzzy disc, the $N$ angle states are defined as
\be
\ket{\varphi_m} =  \frac{1}{\sqrt{N}} \sum_{n=0}^{N-1}
e^{in\varphi_m} \ket{M+n}, 
\ee 
where $\varphi_m = \frac{2\pi}{N}m$ ($m=0, 1, \cdots, N-1$) are the same as before.

\section{Applications} 

\subsection{Angular Noncommutative Solitons in Scalar Field Theory}
\label{2.1}

As the first application of the arguments so far, we would like to consider a scalar field  theory on the fuzzy disc, which is the fuzzy disc version of \cite{Gopakumar:2000zd}.
For a scalar field $\Phi (z,\bar{z})$ defined on the fuzzy disc with radius 
$R=\sqrt{2N\theta}$, its energy functional is given by%
\footnote{
In \cite{Gopakumar:2000zd}, the authors obtained 
this energy functional by taking a limit 
of large noncommutativity; $\theta \to \infty$ 
and rewriting it in the rescaled coordinates. 
The kinetic term, which is usually contained in the action, 
becomes negligible compared with the potential term 
by this operation. 
} 
\be
E =  \int_{D} d^2z \ V_{\st} (\Phi).
\ee
Here the potential $V_\st(\Phi)$ is a polynomial 
\be
V_{\st} (\Phi) = \frac{b_2}{2} \Phi\st \Phi + \frac{b_3}{3}\Phi\st\Phi\st\Phi+\cdots,
\ee
of the field $\Phi$ with respect to the Wick-Voros product, and 
$b_r$'s are constants. 
Note that the field $\Phi$ is also considered to be a 
finite $N$ Hermitian matrix $\hat{\Phi}$ in the number basis via 
the Weyl-Wigner correspondence.
Our argument here is valid even for $N\to\infty$, 
that is, the Moyal plane case.  

The energy functional is extremized by solutions of the following 
equation of motion
\be
0 = \frac{\del V_\st}{\del \Phi} 
=b_2 \Phi + b_3 \Phi\st\Phi + b_4 \Phi\st\Phi\st\Phi + \cdots
\label{scalarEOM1}
\ee
In the commutative case (that is, $\theta = 0$), it admits only constant solutions because of 
the absence of the kinetic term:
$
\Phi (z,\bar{z})= \lambda_*, 
$
where $\lambda_*$ is one of the various extrema of the function $V(x)$, i.e., a real root of the algebraic equation, 
$b_2 x + b_3 x^2 + b_4 x^3 + \cdots = 0$.
On the contrary, for nonzero $\theta$, 
there do exist nontrivial solutions whose energy densities are localized in some regions.
In fact, associated with any projection operator $\hat{e}$ or its symbol $e(z,\bar{z})$, which satisfies $e\st e =e$,   
\be 
\Phi = \lambda_* e(z,\bar{z}), 
\ee
is a solution. 
This is a straightforward application of the argument given in \cite{Gopakumar:2000zd} to the fuzzy disc case.

The point here is that we can choose (sum of) the angular projection 
operators $\hat{\pi}_k$'s.
Namely, 
\be
\Phi = \lambda_* \pi_k (r,\varphi )
=\frac{\lambda_*}{N}\sum_{m,n=0}^{N-1}
e^{-\frac{r^2}{2\theta}} 
\frac{r^{m+n} }{\sqrt{m!n!(2\theta)^{m+n}}}
e^{i(m-n)(\varphi_k -\varphi)},
\ee
solves (\ref{scalarEOM1}) for $k= 0, 1, \cdots, N-1$. 
We call it as an angular soliton solution.
As we have seen, it shares the unit fun-shaped area in the disc.
By using the correspondence
\be
\int_D d^2 z \quad 
\leftrightarrow
\quad 
2\pi\theta\Tr_N, 
\ee
and using that $\hat{\pi}_k$ is rank $1$, it is shown that this solution carries the energy 
\be
E = \int d^2 z V_\st (\lambda_* ) \pi_k (r,\varphi )
= 2\pi\theta V_\st (\lambda_* ).
\ee

On the Moyal plane, the scalar field theory can be regarded as an effective theory for the tachyon field on a non-BPS D$2$-brane, and the solution $\Phi=\lambda_* (1-p_n)$ based on the projection operator $\hat{p}_n$ can be identified as a single D$0$-brane \cite{Dasgupta:2000ft, Harvey:2000jt}, because of the rank of $\hat{p}_n$ is $1$.
In this respect, our angular soliton $\Phi=\lambda_* (1-\pi_k)$ has the same energy as a D$0$-brane, but its shape is completely different. 
Concerning this point, the angular noncommutative solitons 
might be related to the D$0$-brane with orbital angular momentum 
like the optical vortex \cite{Allen:1992zz,YP,UT}. 
This issue will be reported in our forthcoming 
paper \cite{AK3}.



\subsection{Angular Noncommutative Solitons in Gravity}

As the second application, we would like to consider the gravitational system 
we referred in \cite{Asakawa:2009yb}. 
We exploit the first order formulation of a three-dimensional theory of gravity 
on a noncommutative ${\mathbb R}^3$ 
which has a cosmological constant term only, 
\be
\label{theory}
S = -\frac{\Lambda }{\kappa^2}
\int dt d^2z  \ E^{\star},
\ee
where $\Lambda$ is a cosmological constant. 
Here $E^{\star}$ is the $\star$-determinant defined by 
\be
\label{detE}
E^{\star} = \mbox{det}_{\star} E
= \frac{1}{3!}\epsilon^{\mu\nu\rho} \epsilon_{abc} 
E_{\mu}^a  \star E_{\nu}^b  \star E_{\rho}^c,  
\ee
where $E_{\mu}^a (z,\bar{z})$ is a vielbein. 
We denote spacetime indices by $\mu, \nu, \rho$ and 
tangent space indices by $a, b, c$. 
All indices run from $0$ to $2$.  
The metric is also defined through the star product 
in a similar way \cite{Banados:2001xw, Aschieri:2005zs}:
\be 
\label{G}
G_{\mu\nu} = \frac{1}{2}\left(E_{\mu}^a\star E_{\nu}^b 
+ E_{\nu}^b \star E_{\mu}^a\right) \eta_{ab},
\ee
where $\eta_{ab}$ is an $SO(1,2)$ invariant metric of the local Lorentz frame.
We do not assume that $E_{\mu}^a$ or $G_{\mu\nu}$ are invertible as 
$3\times 3$ matrices, that is, we allow degenerate metrics.
From this action, we obtain nine equations of motion 
for ${}^\forall \mu$ and ${}^\forall a$ \cite{Asakawa:2009yb},
\be
\epsilon^{\mu\nu\rho}\epsilon_{abc}\{E_{\nu}^b, E_{\rho}^c\}_{\star}=0,
\label{eom}
\ee
where we used the star-anticommutator defined by
$
\{f,g\}_{\star} \equiv \frac{1}{2}(f\star g+g\star f).
$


In \cite{Asakawa:2009yb}, we gave various kinds of non-trivial solutions for (\ref{eom}),
but all of them are based on the radial projection operators $\hat{p}_n$.
By replacing $\hat{p}_n$ with the angular projection operators $\hat{\pi}_k$, 
various new kinds of solutions 
can be obtained.
We here give an example of them.



The simplest solution is a diagonal vielbein where three 
components are proportional to angular projection operators such as
\be
E_{\mu}^a = \left(
\begin{array}{ccc}
E_0^0  & 0  & 0 \\
0 & E^1_1  & 0  \\
0 & 0  & E_2^2
\end{array}
\right)
=\left(
\begin{array}{ccc}
\alpha_0 \pi_0^{(N)}  & 0  & 0 \\
0 & \alpha_1 \pi_1^{(N)}  & 0  \\
0 & 0  & \alpha_2 \pi_2^{(N)}
\end{array}
\right),
\ee
where $\alpha_0, \alpha_1$ and $\alpha_2$ are arbitrary constants.
Any three different choices among $\pi_0^{(N)}, \pi_1^{(N)}, 
\cdots, \pi_{N-1}^{(N)}$, or three linear combinations of them
can be solutions as long as they are orthogonal among them.  
For the example above with $N=3$, the metric (\ref{G}) is given by
\be
ds^2 = -\alpha_0^2 \pi_0^{(3)} dt^2 
+\alpha_1^2 \pi_1^{(3)}dx^2 
+ \alpha_2^2 \pi_2^{(3)}dy^2, 
\label{diagonal}
\ee
where 
\ba
\pi_k^{(3)}(r, \varphi) 
&=& \frac{1}{3}e^{-r^2/\theta}
\Bigg[
1+\frac{2r}{\theta^{1/2}}\cos(\varphi -\varphi_k^{(3)}) \nonumber \\
&&+\frac{r^2}{\theta}\left\{1+\sqrt{2}\cos [2(\varphi-\varphi_k^{(3)})] \right\}
+\frac{\sqrt{2}r^3}{\theta^{3/2}} \cos(\varphi -\varphi_k^{(3)}) 
+\frac{r^4}{\theta^2}
\Bigg]. 
\label{pi3k}
\ea
Note that $\pi_i^{(N)}$ is idempotent with respect to the Wick-Voros product. 
\begin{figure}[t]
  \begin{center}
   \includegraphics[scale=0.45, bb=0 0 259 214]{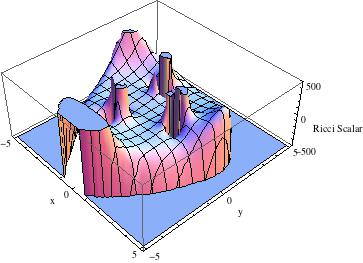}
   \hspace{3cm}
    \includegraphics[scale=0.45, bb=0 0 259 214]{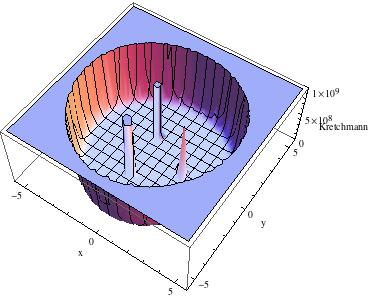}
    \caption{The ordinary (commutative) Ricci scalar 
    and the Kretchmann invariant for the line element (\ref{diagonal}). 
    We set $\alpha_0 = \alpha_1 = \alpha_2 =1$ and $\theta =1$.} 
    \label{N03RicciKret}
  \end{center}
\end{figure}
Considering this metric as an ordinary (that is, commutative) one, one can formally calculate several quantities such as the Riemann tensor.
Figure \ref{N03RicciKret} shows the ordinary Ricci scalar and 
the Kretchmann invariant 
for the metric (\ref{diagonal}), respectively. 
Both quantities are almost flat in the disc region but diverge outside the disc, which suggest that the spacetime corresponding to this solution is the fuzzy disc with radius $R=\sqrt{6\theta}$. 
This matches to the fact that $R=\sqrt{2N\theta}$ and $N=3$ here. 
The two invariants also diverge around three points, where 
$\pi_0^{(3)}, \pi_1^{(3)}$ and $\pi_2^{(3)}$ have their peaks.
These divergences would be artifacts due to the bad choices 
of the observables, and would be resolved if we define more admissible quantities. 
We emphasize that the fuzzy disc is an emergent space in this model, that is, its size $N$ is not a parameter of the theory but a parameter of a solution.

\section{Conclusion and Discussion}

In this paper, we investigated the fuzzy disc
by introducing the concept of angles.
By defining the angle states $\ket{\varphi_m}$ and the angle operator $\hat{\varphi}$, which are known as the phase states and the phase operator in quantum optics,
we reformulate the $N\times N$ matrix algebra for the fuzzy disc as the commutation relations among 
the operators $\hat{\varphi}$, $\hat{N}$, $\hat{U}$ and $\hat{V}$.
The angle states were also used to divide the fuzzy disc into fan-shaped regions. 
This type of division of the fuzzy disc has not been 
considered so far.

As an application, 
we found noncommutative angular solitons 
for a scalar field theory on the fuzzy disc,
where fan-shaped regions might be  related to D$0$-branes.
Though it needs further study to support on this point, 
it might be possible to identify the angular noncommutative solitons 
with the vortex motion of D$0$-branes 
like the optical vortex in laser physics.  
The extensions of angular solitons to exact angular solitons 
and multi-angular solitons would be possible.

As a generalization, we also described briefly the fuzzy annulus 
and the angle states for it.
If there are orthonormal $N$ states, one can construct the angle states.  
Because of this, there is a lot of 
possibility of applications of these kinds of constructions for fuzzy objects. 
For example, if we choose two sets of states, $\{\ket{0}, \ket{1}, \cdots, \ket{N-1}\}$ and 
$\ket{M}, \ket{M+1}, \cdots, \ket{M+N-1}$ with $N<M$, we obtain a disjoint union 
of the fuzzy disc and a fuzzy annulus. 
One can also constitute two sets of the $N$ angle states. 
Because two sets are orthogonal with each other, 
they form $2N$ orthogonal angle states.

In particular, the application of the fuzzy disc or other fuzzy objects  
to investigate black hole microstates 
would be interesting, because the feature of holography is encoded in this setting.
To this end, we would have to understand the edge states 
in this setup further. 

As a concluding remark, we refer a possibility to relate 
noncommutative theories to experiments. 
The solutions of a scalar field theory in a noncommutative 
space, e.g., the GMS solitons, are written in terms of the function
known as the Laguerre-Gaussian function in quantum optics and 
laser physics \cite{Allen:1992zz,YP,UT}.
The Laguerre-Gaussian function is the solution of 
the Helmholtz equation with cylindrical symmetry in the three-dimensional 
space. One can actually make the laser beam expressed by this kind
of function with the Gaussian profile radially to the direction 
along a beam goes. The profile along this direction is arbitrary 
and the point where the spread of the beam is narrowest 
is called ``the waist" from its shape.
The size of this waist can be identified to the magnitude of the 
noncommutativity parameter $\theta$ when we assign the noncommutative 
solitons to the cross section of the Laguerre-Gaussian beam. 
The intensity of the beam seems to correspond to 
the height of the noncommutative solitons at the level of equations. 
Although the Laguarre-Gaussian beam is not 
a ``real" noncommutative soliton but 
an analogue to it, 
we might be able to 
say something about quantum gravity by certain experiments with laser,
because a noncommutative theory could be 
an effective theory of quantum gravity. 
If this idea is somehow justified, it would be a very exciting issue.

\section*{Acknowledgements}
The authors would like to thank Y. Kurita, F. Lizzi and H. Saida
for helpful comments and discussions. 
This work of S. K. is supported 
by JSPS Grand-in-Aid 
for Young Scientists (B) 21740198. 









\bibliographystyle{JHEP}
\bibliography{refsAK2}

\end{document}